# Interactive Web Application for Exploring Matrices of Neural Connectivity

David J. Caldwell[†], *Student Member, IEEE*, Jing Wu[†], *Student Member, IEEE*, Kaitlyn Casimo, Jeffrey G. Ojemann, Rajesh P.N. Rao, *Member, IEEE*

*Abstract*— We present here a browser-based application for visualizing patterns of connectivity in 3D stacked data matrices with large numbers of pairwise relations. Visualizing a connectivity matrix, looking for trends and patterns, and dynamically manipulating these values is a challenge for scientists from diverse fields, including neuroscience and genomics. In particular, high-dimensional neural data include those acquired via electroencephalography (EEG), electrocorticography (ECoG), magnetoencephalography (MEG), and functional MRI. Neural connectivity data contains multivariate attributes for each edge between different brain regions, which motivated our lightweight, open source, easy-to-use visualization tool for the exploration of these connectivity matrices to highlight connections of interest. Here we present a client-side, mobile-compatible visualization tool written entirely in HTML5/JavaScript that allows in-browser manipulation of user-defined files for exploration of brain connectivity. Visualizations can highlight different aspects of the data simultaneously across different dimensions. Input files are in JSON format, and custom Python scripts have been written to parse MATLAB or Python data files into JSON-loadable format. We demonstrate the analysis of connectivity data acquired via human ECoG recordings as a domain-specific implementation of our application. We envision applications for this interactive tool in fields seeking to visualize pairwise connectivity.

## I. INTRODUCTION

In this paper, we present an interactive exploration tool specifically tailored to visualize large numbers of pairwise neural connectivity dynamics. We further demonstrate the utility of this tool for analyzing pairwise connections across various regions of the brain based on electrocorticography (ECoG) data.

Interpreting neural connectivity is a challenging problem because cortical connectivity visualizations are largely exploratory and require the encoding of many multivariate attributes that exist for each pairwise edge, as each edge can be its own high-dimensional correlation matrix. These multiple metrics for each connection require an investment of user time to navigate and interpret. Unlike social networks or aggregate flow for which directed graphs are often designed, neural connectivity is highly nonstationary through time, and the set of descriptive statistics change rapidly. This adds many potential encoding dimensions for any visualization tool, leading to a critical need for user interactivity to explore data and formulate appropriate questions. Fig. 1 highlights sources of data, and potential dimensions along which a user may seek to subselect and visualize the data.

[†] These authors contributed equally

*Research supported by grants from NSF EEC-1028725, NINDS 5R01NS065186, the UW Big Data for Genomics and Neuroscience (BDGN) T32 Training Grant, and the WRF Fund for Innovation in Neuroengineering.

D.J Caldwell is with the department of Bioengineering, University of Washington, Seattle, WA 98105 USA, and the Center for Sensorimotor Neural Engineering (CSNE). (phone: 206-685-3301; e-mail: djcald@uw.edu).

J. Wu is with the department of Bioengineering, University of Washington, Seattle, WA 98105 USA, and the CSNE. (phone: 206-685-3301; e-mail: jiwu@uw.edu).

K. Casimo is with the Graduate Program in Neuroscience, University of Washington, Seattle, WA 98105 USA, and the CSNE (e-mail: kcasimo@uw.edu)

J.G. Ojemann is with the department of Neurological Surgery, University of Washington, Seattle, WA 98105 USA, and the CSNE. (e-mail: jojemann@uw.edu).

R.P.N. Rao is with the department of Computer Science, University of Washington, Seattle, WA 98105 USA, and the CSNE. (e-mail: rao@cs.washington.edu).

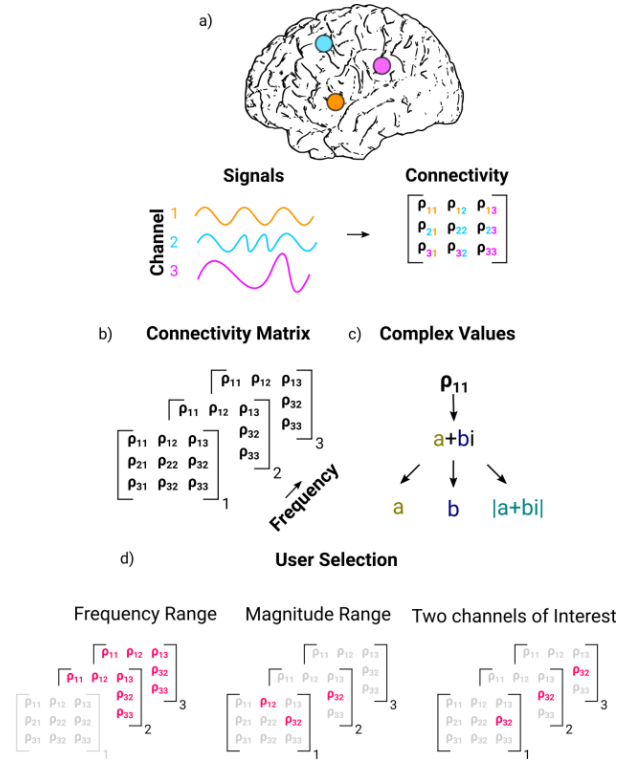

Figure 1. a) Sources of ECoG signals, which can be generalized to any spatially specific measurement of correlation. From these time series traces, researchers calculate matrices of connectivity values. b) Connectivity matrices often exist for various frequency bands of interest, and c) are often complex valued. d) Various dimensions of matrix selection.

Our dataset explorations emphasize the use of the tool to analyze pairwise connectivity in ECoG datasets. The utilization of our tool demonstrates a unique opportunity to apply and extend existing interactive visualization techniques, such as chord diagrams, dynamic querying and selection, and interactivity, to exploratory neuroscience.

## II. RELATED WORK AND MOTIVATION

Visualizing neural connectivity is a significant problem in the field of neuroscience. Examples of current tools to address this problem can be found in a centralized repository named "The Neuroimaging Informatics Tools and Resources Clearinghouse" (NITRC) [1] founded by the National Institutes of Health (NIH), which hosts hundreds of tools (mostly for MRI data) for the analysis and visualization of Neuroimaging data. A subset of tools for MEG and ECoG data are available as MATLAB plugins, which require package download, a MATLAB license, and a user interface with a powerful but complicated software suite. We propose instead an open source, browser-based visualization toolkit.

Other past visualization techniques include rendering each anatomical location on a virtual cortex. However, this approach often leads to occlusions in the resulting visualization due to the sheer number of overlapping connection pairs. Additional techniques include static graphs of frequency dependent functional connectivity through undirected graphs based on fMRI data [2]. Of note is the labeling of various anatomic locations in free space, with a web of connections. While this common visualization demonstrates connectivity between numerous nodes and captures the connectivity of the system, we sought a more intuitive way to represent connectivity.

A review of various methods [3] demonstrates different approaches used in the past for demonstrating both functional connectivity (correlations in activity between various regions) and anatomic connectivity (fiber tract density between regions). Visualization strategies range from literal demonstration of the brain's anatomical structure, to more abstract visualizations depicting the brain with nodes and edges. We sought for our application a simple, uncluttered interface highlighting connectivity patterns without requiring rotation to visualize obscured regions.

Other recent advances in visualization of neural data include the folding of time curves to visualize patterns of temporal activation [4] and stacking of adjacency matrices representing temporal snapshots to visualize changes in networks over time [5]. While these tools offer great flexibility and exploratory depth in dealing with time series data, we desired a simpler, more lightweight approach to visualize connectivity matrices which were constant across a time period. We wished to implement strategies similar to the interactive pruning [5] and rapid exploratory nature of the above tools. We also desired a tool agnostic to the underlying data matrix, enabling the use of whichever connectivity value the researcher desires.

## III. METHODS AND IMPLEMENTATION

### A. Visualization Architecture

Our visualization architecture, with corresponding static screenshot, is shown in the schematic in Fig. 2a. A correlation matrix (3D matrix of connectivity strengths at various sensor locations and frequency ranges) is loaded into a client-side JavaScript-enabled page in a desktop or mobile browser via drag and drop or a browsing menu. The correlation matrices, nested arrays in JavaScript Object Notation (JSON) format, are loaded using HTML5 standards compatible file handing routines. These datasets can be generated by the end-user, by conversion from either a Python numpy data file or a MATLAB data file, using our included Python script. These two data formats represent the vast majority of pairwise connectivity data in neuroscience.

The basis upon which our visualization builds upon is the d3.js [6] implementation of Krzywinski's "circos" [7], which we chose as a visualization technique for representing pairwise connections with minimal occlusion. Our implementation can handle non-symmetric connectivity matrices, where there is directional connectivity strengths between nodes. For additional visualization dimensions, we simultaneously plot dynamic bar charts between pairs of electrodes across frequency bands, as well as a global histogram that demonstrates the average strength across all frequencies for all nodes. For the implementation to be functionally usable, the interface must be extended with interactivity that is rapid and responsive when selecting and visualizing more information about pairwise connections.

The user selects a frequency range of interest, a statistical measure of interest (absolute value or magnitude and phase information, if both are present), the inclusion or exclusion of self-connections, and then presses the "re-slice matrix and render" button to submit the query and update the chord diagram (Fig. 2b). This uses the JavaScript math.js library to subselect and compute an average matrix slice across the selected frequencies. The browser-based controls use JavaScript d3.js and jQuery controller frameworks to create cross-platform, responsive controls. The user-selected threshold cutoff value dynamically prunes connections above or below the selected strength from the diagram in real time. A callback with active tweening of the arc and chord shapes with d3 transitions conveys a smooth but rapid transition between rendered frames (Fig. 3a).

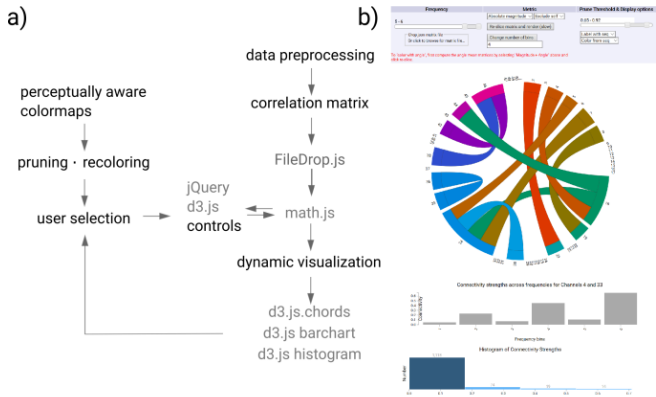

Figure 2. a) Schematic overview of the approach to data visualization. b) Example screenshot from the visualization highlighting the use of subselection, thresholding, dynamic bar graph generation, and hover feature.

As the color representation of nominal categories is extremely important [8] and spatial information must be accurately conveyed, we tailored a CIE L*a*b colorscale that fades in two directions, as well as an L*c*h colorscale that transitions in color unidirectionally (Fig. 3b). In this way, the two CIE L*a*b colors correspond to spatial locations on our ECoG grid. For datasets where spatial ordering is not as clear, the L*c*h unidirectional colorscale emphasizes separability of nodes. Additional color files can be supplied by the user to tailor the data conveyed by color.

*B. User Interactions*

Once the chord diagram is set up for visualization, the user hovers over a node or a chord to display connection strengths for that node, or between the two nodes specified by a given chord. An additional user option for coloring via mean phase angle of the connection uses a divergent color palette from ColorBrewer (colorbrewer.org) to show differences between positive and negative phase angles in complex data.

Upon clicking a chord, a bar graph dynamically updates to display connection strengths across frequencies for the two nodes of interest (Fig. 2b). By hovering over any bar in the chart, the connectivity strength for that frequency will be displayed. These features were implemented using the "hover" and "on click" functionalities from JS and d3 (Fig. 3c, 3d). By revealing the connection strengths across all frequencies, the user gains additional insight into the topology of connectivity for two regions without losing the global connectivity pattern provided by the chord diagram.

For any selected set of frequencies, a histogram displays all channel-by-channel connectivity values in a given frequency range (Fig. 3b). Using interactive JS functionality, users input a desired bin size to visualize different binnings of their data, allowing for rapid coarse and fine visualizations. By the combination of dynamic features, and three separate visualization strategies (chord diagram, bar chart, and histogram), for the data, users can interpret their datasets in a multifaceted manner without any software downloads or programming.

Figure 3. a) Example screenshot illustrating rapid fade and tween pruning using a single slider to select for maximally salient chords and remove occlusion. Note that regions with similar colors convey a local group of intra-region connections while dissimilar colors convey longer distance, crosscortical connections. The user then has the option to color the chord diagram based off anatomic locations from a file (if present), or in a constant luminance scheme based off sequential order. b) A modified L*a*b colorspace controlled for luminance for coloring arcs representing spatial locations around the chord diagram to give the user rapid assessment of sensor proximity. Hover-over functionality for both c) the bar chart and d) chord diagram.

*C. Sample analysis data*

All provided datasets were acquired with written informed consent through a protocol approved by the University of Washington Institutional Review Board. Our sample data sets are human ECoG datasets acquired during a resting state, where the subject was not actively engaged in any task. Data were processed for connectivity metrics [9].

IV. RESULTS AND DISCUSSION

The screenshot in Fig. 2b illustrates our visualization tool. We use this tool to explore connectivity matrices of differing values, including phase locking value (PLV) [9], phase slope index (PSI), correlation, and coherence, as well as intraclass correlation (ICC) of these measures over time [10]. Of note is the ability of the application to handle different statistical measures of connectivity, highlighting its extensibility to other data types and data sets. Of note is that our example involves symmetric connectivity, but our application can handle non-symmetric matrices. In this instance, the chord would be thicker near the node with the greater value for the asymmetric matrix. This is useful when looking at connections that are stronger in one causal direction than another.

One particular application highlights (Fig. 4) the use of our tool to explore ICC over three sessions of PLV in the high gamma band (HG, 70-200 Hz in ECoG) between various regions of the cortex across 9 subjects. PLV describes the consistency of a phase difference between two signals throughout time, while ICC is an ANOVA-based metric for the evaluation of the stability of connections over time across multiple individuals. The stationarity of brain signals is a topic of research interest in assessing the function and structure of cortical networks. HG signals in ECoG are known to be associated with local synchronous neural activity [11]. We use our application to first prune high gamma (70-200 Hz in frequency) ICC values to a low ICC value range (Fig. 4a), revealing widespread unstable connections between regions. Although difficult to currently interpret, this visualization allows us to further analyze these highly unstable regions. Further exploration of high ICC values demonstrates fewer, more local stable connections over time (Fig. 4b). An example chord of interest illustrates the high ICC between Brodmann areas regions 4 and 34, representing a highly stable connection between motor and entorhinal cortex. Entorhinal cortex is associated with spatial processing [12], positing interesting questions to explore as to why this region might have stable connections with motor cortex. This finding highlights the use of our tool to explore novel connectivity patterns uniquely available from human electrophysiological recordings.

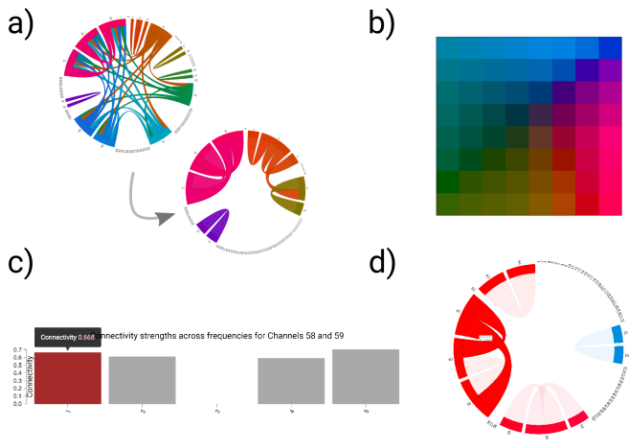

A different data set explored by our research group assesses the spatial representation of connections across the cortex. Using PLV as a metric, we first subselect and visualize connectivity related to the beta band (13-30 Hz, Fig. 5a) and PLV between 0.65 and 1. After switching to high frequency (high gamma) PLV and thresholding for the same values (Fig. 5b), we observe a different pattern of connectivity. This emphasizes the utility of our tool as an exploratory form of analysis to inform further inquiries.

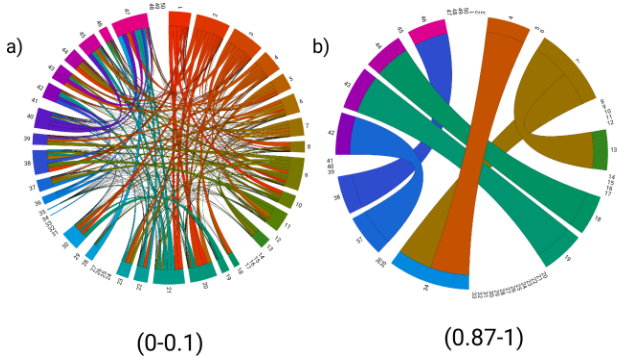

Figure 4. ICC values in the high gamma (70-200 Hz in ECoG) range for nine subjects. a) Low thresholded values, illustrating many long-range unstable connections. b) High thresholded values, illustrating fewer, mostly local, stable connections. We highlight the connection between Brodmann areas 4 and 34, motor cortex and entorhinal cortex, respectively.

Our results for test matrices of size 65x64x64 and 5x64x64 reveal re-rendering and `math.js` processing within 500 ms to a second depending on the user's' machine. Code profiling revealed the bottlenecks to be in the `math.js` processing and computations on matrices, so in the event of matrices orders of magnitude larger than our current ones, we may need to consider alternative approaches. Since the visualization tool does not require any additional software packages, this is a significant advantage over existing tools that may require MATLAB or other software on the end user's computer.

In conclusion, we demonstrate a browser-based, open source, easy to use application to quickly visualize, prune, reslice and explore complex data matrices encoding relationships between pairwise nodes.

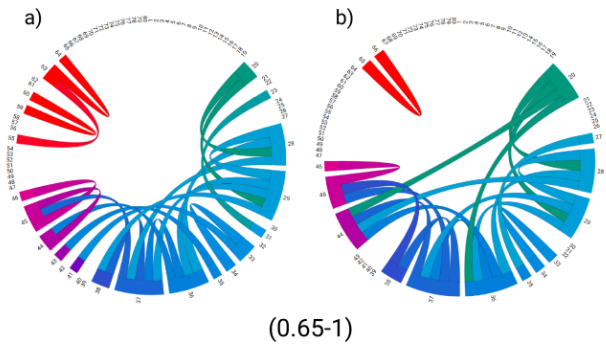

Figure 5. PLV values for a single subject. a) Beta Band (13-30 Hz) displaying local, looping connections. b) High gamma, 70-200 Hz), displaying a different pattern of connectivity.

ACKNOWLEDGMENT

The authors acknowledge Dr. Emily Fox, Dr. Nick Foti, and Dr. Jeffrey Heer for conversation and feedback. The authors would also like to thank all the patients who willingly contributed to the data presented in this paper.

DISCLAIMER

The views expressed are solely those of the authors and do not reflect the views of the National Institutes of Health or the National Science Foundation.

SOFTWARE LINKS

Overview - http://uwgridlab.github.io/Affinity/

Software - http://uwgridlab.github.io/Affinity/main.html